\documentclass[12pt]{article}
\usepackage[latin1]{inputenc}
\usepackage[british]{babel}
\usepackage{cmap}
\usepackage{lmodern}

\usepackage{amssymb, amsmath, amsthm}
\usepackage[a4paper,top=25mm,bottom=25mm,left=25mm,right=25mm]{geometry}
\usepackage{etex}
\usepackage{ragged2e}

\usepackage{authblk} 
\usepackage{pifont}
\usepackage{graphicx}
\usepackage[usenames,dvipsnames,svgnames,table]{xcolor}
\usepackage[figuresright]{rotating}
\usepackage{xtab} 
\usepackage{longtable} 
\usepackage{multirow}
\usepackage{footnote}
\usepackage[stable]{footmisc}
\usepackage{chngpage} 
\usepackage{pdflscape} 
\usepackage[nottoc,notlot,notlof]{tocbibind} 

\usepackage{pgfplots}
\pgfplotsset{compat=1.14}
\pgfplotsset{every tick label/.append style={font=\footnotesize}}
\usepackage{setspace}

\makesavenoteenv{tabular}
\usepackage{tabularx}
\usepackage{booktabs}
\usepackage{threeparttable} 
\usepackage[referable]{threeparttablex} 
\newcolumntype{R}{>{\raggedleft\arraybackslash}X}
\newcolumntype{L}{>{\raggedright\arraybackslash}X}
\newcolumntype{C}{>{\centering\arraybackslash}X}
\newcolumntype{A}{>{\columncolor{gray!25}}C}
\newcolumntype{a}{>{\columncolor{gray!25}}c}

\newlength{\tablen}

\usepackage{dcolumn} 
\newcolumntype{.}{D{.}{.}{-1}}

\usepackage{tikz}
\usetikzlibrary{arrows, calc, matrix, patterns, positioning, trees}
\usepackage[semicolon]{natbib}
\usepackage[hyphens]{url}
\usepackage{hyperref} 
\hypersetup{
  colorlinks   = true,    
  urlcolor     = blue,    
  linkcolor    = blue,    
  citecolor    = ForestGreen      
}
\usepackage{microtype}
\usepackage[justification=centering]{caption} 

\usepackage[labelformat=simple]{subcaption}

\DeclareCaptionLabelFormat{parenthesis}{(#2)}
\makeatletter
\renewcommand\p@subfigure{\arabic{figure}.}
\makeatother

\DeclareCaptionLabelFormat{parenthesis}{(#2)}
\makeatletter
\renewcommand\p@subtable{\arabic{table}.}
\makeatother

\usepackage{enumitem}
\setlist[itemize]{leftmargin=2.5\parindent}
\setlist[enumerate]{leftmargin=2.5\parindent}

\newenvironment{customlegend}[1][]{%
	\begingroup
	\csname pgfplots@init@cleared@structures\endcsname
	\pgfplotsset{#1}%
    }{%
	\csname pgfplots@createlegend\endcsname
	\endgroup
    }%
\def\addlegendimage{\csname pgfplots@addlegendimage\endcsname}

\theoremstyle{plain}

\theoremstyle{definition}

\theoremstyle{remark}

\def\keywords{\vspace{.5em} 
{\noindent \textit{Keywords}: }}

\def\JEL{\vspace{.5em} 
{\noindent \textbf{\emph{JEL} classification number}: }}

\def\AMS{\vspace{.5em} 
{\noindent \textbf{\emph{MSC} class}: }}

\author{\href{https://sites.google.com/view/laszlocsato}{L\'aszl\'o Csat\'o}\thanks{~E-mail: \emph{laszlo.csato@sztaki.hu}} }
\affil{Institute for Computer Science and Control (SZTAKI) \\
E\"otv\"os Lor\'and Research Network (ELKH) \\
Laboratory on Engineering and Management Intelligence \\
Research Group of Operations Research and Decision Systems}
\affil{Corvinus University of Budapest (BCE) \\
Department of Operations Research and Actuarial Sciences}
\affil{Budapest, Hungary}
\title{Fair tournament design: A flaw of \\ the UEFA Euro 2020 qualification}
\date{\today}

\def\Dedication{
{\noindent
$\mathfrak{Wer}$, $\mathfrak{von}$ $\mathfrak{inneren}$ $\mathfrak{Kr\ddot{a}ften}$ $\mathfrak{angeregt}$, $\mathfrak{sich}$ $\mathfrak{ein}$ $\mathfrak{solches}$ $\mathfrak{Werk}$ $\mathfrak{vorsetzen}$ $\mathfrak{will}$, $\mathfrak{der}$ $\mathfrak{r\ddot{u}ste}$ $\mathfrak{sich}$ $\mathfrak{zu}$ $\mathfrak{dem}$ $\mathfrak{frommen}$ $\mathfrak{Unternehmen}$ $\mathfrak{mit}$ $\mathfrak{Kr\ddot{a}ften}$ $\mathfrak{wie}$ $\mathfrak{zu}$ $\mathfrak{einer}$ $\mathfrak{weiten}$ $\mathfrak{Pilgerfahrt}$ $\mathfrak{aus}$.
$\mathfrak{Er}$ $\mathfrak{opfere}$ $\mathfrak{Zeit}$ $\mathfrak{und}$ $\mathfrak{scheue}$ $\mathfrak{keine}$ 
$\mathfrak{Anstrengung}$, $\mathfrak{er}$ $\mathfrak{f\ddot{u}rchte}$ $\mathfrak{keine}$ $\mathfrak{zeitliche}$ $\mathfrak{Gewalt}$ $\mathfrak{und}$ $\mathfrak{Gr\ddot{o} \ss e}$, $\mathfrak{er}$ $\mathfrak{erhebe}$ $\mathfrak{sich}$ $\mathfrak{\ddot{u}ber}$ $\mathfrak{eigene}$ $\mathfrak{Eitelkeit}$ $\mathfrak{und}$ $\mathfrak{falsche}$ $\mathfrak{Scham}$, $\mathfrak{um}$ $\mathfrak{nach}$ $\mathfrak{dem}$ $\mathfrak{Ausdruck}$ $\mathfrak{des}$ $\mathfrak{franz\ddot{o}sischen}$ $\mathfrak{Kodex}$
\textcolor{ForestGreen}{$\mathfrak{die}$ $\mathfrak{Wahrheit}$ $\mathfrak{zu}$ $\mathfrak{sagen}$, $\mathfrak{nichts}$ $\mathfrak{als}$ $\mathfrak{die}$ $\mathfrak{Wahrheit,}$ $\mathfrak{die}$ $\mathfrak{ganze}$ $\mathfrak{Wahrheit}$}.\footnote{~
``Whoever, stirred by ambition, undertakes such a task, let him prepare himself for his pious undertaking as for a long pilgrimage; let him give up his time, spare no sacrifice, fear no temporal rank or power, and rise above all feelings of personal vanity, of false shame, in order, according to the French code, to speak \emph{the Truth, the whole Truth, and nothing but the Truth}.'' (Source: Carl von Clausewitz: \emph{On War}, Book 2, Chapter 6 -- On Examples. Translated by Colonel James John Graham, London, N. Tr\"ubner, 1873. \url{http://clausewitz.com/readings/OnWar1873/TOC.htm})}
}
\vspace{0.25cm}

\flushright
\noindent (Carl von Clausewitz: \emph{Vom Kriege})

\vspace{1cm} 
\justify }

\begin{document}

\newgeometry{top=5mm,bottom=10mm,left=20mm,right=20mm}

\maketitle
\thispagestyle{empty}
\Dedication

\begin{abstract}
\noindent
The integrity of a sport can be seriously undermined if its rules punish winning as this creates incentives for strategic manipulation. Therefore, a sports tournament can be called unfair if the overall win probabilities are not ordered according to the teams' ranking based on their past performances. We present how statistical methods can contribute to choosing a tournament format that is in line with the above axiom. In particular, the qualification for the 2020 UEFA European Championship is shown to violate this requirement: being a top team in the lowest-ranked League D of the 2018/19 UEFA Nations League substantially increases the probability of qualifying compared to being a bottom team in the higher-ranked League C. The unfairness can be remarkably reduced or even eliminated with slightly changing the path formation policy of the UEFA Euro 2020 qualifying play-offs. The misaligned design has severely punished a team for winning a match years before. Since the deficiency is an inherent feature of the qualifying process, the Union of European Football Associations (UEFA) should reconsider the format of future tournaments to eliminate the unfair advantage enjoyed by certain teams.



\keywords{fairness; mechanism design; simulation; soccer; UEFA Euro 2020}

\AMS{62F07, 68U20}

\JEL{C44, C63, Z20}
\end{abstract}

\clearpage
\restoregeometry

\section{Introduction} \label{Sec1}

Fairness in sports is an increasingly discussed problem in the academic literature, concerning various topics such as penalty shootouts \citep{ApesteguiaPalacios-Huerta2010, KocherLenzSutter2012, Palacios-Huerta2014, BramsIsmail2018}, referee assignment \citep{YavuzInanFiglali2008}, scheduling \citep{ChaterArrondelGayantLaslier2021, DuranGuajardoSaure2017, Guyon2020a, KrumerLechner2017, KrumerMegidishSela2020a}, seeding \citep{CeaDuranGuajardoSureSiebertZamorano2020, Csato2020c, Guyon2015a, LalienaLopez2019}, team selection \citep{Harville2003}, or tournament design \citep{ArlegiDimitrov2020, Guyon2018a}. Nevertheless, since any sporting contest establishes a hierarchy among the competitors by differentiating winners and losers, perhaps the most important criterion of the fairness of a sports competition is the well-known Aristotelian Justice principle: the winning probabilities should be ordered according to the players' ranking.
If this property does not hold for a given tournament, players have perverse incentives to manipulate their ranking \citep{GrohMoldovanuSelaSunde2012}, which means a clear design failure \citep{Szymanski2003} that can be identified and, ideally, improved by scientific research.

This paper aims to demonstrate how statistical methods can contribute to choosing an appropriate tournament format.
Similar techniques have been successfully used to compare the abilities of players from different eras in professional sports \citep{BerryReeseLarkey1999}, to model player networks in team sports \citep{HorraceJungSanders2020}, to forecast basketball results by quantile regression \citep{KoenkerBassett2010}, or to predict ultimate world records in athletics \citep{EinmahlMagnus2008}, among others. However, these applications usually require extensive and reliable real datasets, which are not necessarily available if the planned tournament is designed in a novel way. Furthermore, most tournament formats exhibit a high degree of complexity \citep{Guyon2018a}, thus analytical proofs are often impossible to obtain. Unsurprisingly, computer simulations mean a standard approach in the evaluation of different tournament designs \citep{Appleton1995, Csato2019g, DagaevRudyak2019, GoossensBelienSpieksma2012, ScarfYusofBilbao2009}.

On the other hand, it is advised to keep the underlying statistical model as simple as possible because it might help that administrators embrace the results and finally correct the problematic rule. While using a sophisticated methodology can persuade the reviewers of a prestigious journal, its application might even decrease the impact on decision makers who may dismiss the intended message as a mere scientific curiosity without clear real-world implications.

Therefore, we scrutinize a competition design in soccer, probably the most popular sport in the world, from the perspective of fairness, without digging deep into the question of how to predict match results.
In particular, the qualification for the 2020 UEFA European Championship is verified to robustly violate a crucial principle in an elementary probabilistic model: some national teams that are ranked lower at the beginning of the qualifying process have a considerably higher chance to advance to the final tournament than certain higher-ranked teams. The problem is revealed to be caused by the misaligned integration of the 2020 UEFA European Championship qualifying tournament with the inaugural season of a new competition, the 2018/19 UEFA Nations League.

The root of the shortcoming resides in a novel policy of the Union of European Football Associations (UEFA) that aims to increase the diversity of the teams playing in the 2020 UEFA European Championship.
In this sense, the current work illuminates how the promotion of small associations can cause unfairness without a careful analysis in advance.
Similar ill-devised handicapping systems sometimes make losing a profitable strategy, for instance, in sports using player drafts with the traditional set-up of reverse order when a team might decrease efforts to win after it has no more chance to progress in order to obtain an advantage in the player draft \citep{TaylorTrogdon2002, PriceSoebbingBerriHumphreys2010}. \citet{Fornwagner2019} is probably the first study verifying that teams apply a concrete losing strategy. There are several policy proposals to remedy this problem \citep{BanchioMunro2020, Lenten2016, LentenSmithBoys2018, KazachkovVardi2020}.
Analogously, lower handicappers in golf have a higher probability of winning in both stroke play and match play games \citep{McHale2010}.

Such opportunities for manipulation may emerge in soccer, too: \citet{LasekSzlavikGagolewskiBhulai2016} discuss some strategies to improve a team's position in the ``old'' official FIFA ranking, which has been used before the reform in June 2018. The situation can be even worse if a team is strictly better off by losing, not only in expected terms \citep{DagaevSonin2018, Csato2020f, Csato2021a}.
Hence, together with many historical cases when changing sports rules led to unforeseen consequences \citep{KendallLenten2017}, our paper warns the benefit of consultations between sports administrators and the academic community.

The current work is also closely connected to the papers which explicitly check the condition of whether the order of win probabilities coincides with the contestants' ranking or not, albeit without a simulation methodology.
\citet{Hwang1982} shows that this axiom may not hold under the traditional seeding method, but reseeding after each round restores monotonicity.
According to \citet{BaumannMathesonHowe2010}, the 10th and 11th seeds from the regular season average more wins and statistically progress farther than the 8th and 9th teams due to the lack of reseeding in the National Collegiate Athletic Association (NCAA) men's basketball ``March Madness'' tournament. The same issue is investigated by \citet{MorrisBokhari2012}.
\citet{GrohMoldovanuSelaSunde2012} study an elimination tournament with four players and identify all seedings ensuring a higher winning probability to higher-ranked players.

Our main contributions can be summarised as follows:
(1) we offer the first detailed and rigorous documentation of unfairness in the qualification for the 2020 UEFA European Championship;
(2) we provide straightforward solutions to substantially mitigate the problem of perverse incentives such that the UEFA Nations League does not become completely uninteresting;
(3) we present how the misaligned rules have severely punished a team for winning a match years before.
It is worth emphasizing that the most challenging part of the simulation is coding the algorithm of the qualifying play-offs appropriately since the relevant regulation describes only the principles of the team selection and path formation rules and even contains a contradiction \citep{Csato2020g}. Before underrating this achievement, the reader is encouraged to try to understand the corresponding UEFA Media Briefing, which discusses three possible scenarios over 89 slides \citep{UEFA2017g}.

Admittedly, the strange tournament format has already been criticized for allowing weaker teams to qualify through the Nations League in the media \citep{Dunbar2017}.
A Romanian computer programmer called \emph{Eduard Ranghiuc} has provided preliminary calculations to show that losing could improve the chances of participating in the 2020 UEFA European Championship \citep{Ranghiuc2017}. Even though his post reports the same qualitative findings as our study, the details of the methodology are missing and the calculations cannot be reproduced. Therefore, it has severe shortcomings from a scientific point of view but is dutifully acknowledged as a great source of inspiration.

At first sight, the presented example seems to be only an instructive case study for football fans. However, as \citet[p.~1]{Wright2014} writes, ``\emph{given that sports are of great interest to a high percentage of the world's population, it could be counter-argued that there is little that could be researched into that is more important}''.
Our investigation may inspire similar statistical analyses on the fairness of tournament formats and potentially can have an impact on decision makers. There is a recent work presenting the qualification for the 2020 UEFA European Championship as an example of poor tournament design \citep{HaugenKrumer2021}, where an entire subsection (Section~3.2) is devoted to discussing our results.

The paper proceeds as follows. 
The qualification for the 2020 UEFA European Championship is outlined in Section~\ref{Sec2} and the simulation technique is described in Section~\ref{Sec3}.
Section~\ref{Sec4} presents our quantitative results with sensitivity analysis and Section~\ref{Sec5} proposes two solutions to reduce the unfairness of the qualification. Section~\ref{Sec6} studies to what extent the deficiency of the qualifying process depends on the assumed distribution of teams' strength, while Section~\ref{Sec7} turns to some implications for national teams.
The paper ends with concluding remarks in Section~\ref{Sec8}.

\section{Qualification for the UEFA Euro 2020} \label{Sec2}

The \href{https://en.wikipedia.org/wiki/UEFA_Euro_2020}{2020 UEFA European Football Championship}, or shortly the UEFA Euro 2020, is the 16th international men's football championship of Europe. For the first time, it will be spread over 12 cities in 12 countries across the continent, and no national association receives an automatic qualifying berth as a host country.
Although similarly to the UEFA Euro 2016, 24 teams participate in the final tournament, the qualification is fundamentally different from the previous editions because it is strongly connected to the inaugural season of the new competition called the \href{https://en.wikipedia.org/wiki/UEFA_Nations_League}{UEFA Nations League}.

The whole process of qualification for the UEFA Euro 2020 is regulated by two official documents \citep{UEFA2018c, UEFA2018e}.
It starts with the \href{https://en.wikipedia.org/wiki/2018\%E2\%80\%9319_UEFA_Nations_League}{2018/19 UEFA Nations League}. First, the 55 UEFA national teams are divided into four divisions called \emph{leagues}.
In particular, they are ordered according to their UEFA national team coefficients at the end of the 2018 FIFA World Cup qualifiers without the play-offs, and the 12 highest-ranked teams form League A, the next 12 form League B, the next 15 form League C, and the remaining 16 form League D.
The leagues are divided into four groups of three (Leagues A and B), four groups of four (League D), and three groups of four plus one group of three teams (League C) according to the traditional seeding regime: the best four teams are placed into Pot 1, the second best four teams are placed into Pot 2, the next four into Pot 3, the remaining teams (if any) into Pot 4, and one team from each pot is given to each group. The groups are organized as a home-away (double) round-robin tournament, therefore each team plays four or six matches.

After ranking the teams in each group, four league rankings are established, where the four group winners are the first four, the four runners-up are the next four, and so on. Teams having the same position are ranked on the basis of the number of points and goal differences (in the case of League C, without the results against the fourth-placed teams because there is a group with only three teams), except for positions 1--4 that are allocated among the group winners of League A through the \href{https://en.wikipedia.org/wiki/2019_UEFA_Nations_League_Finals}{2019 UEFA Nations League Finals}. However, this event does not affect the qualification for the UEFA Euro 2020.
The league rankings are aggregated into the overall UEFA Nations League ranking such that the 12 teams of League A occupy positions 1--12 according to the ranking in League A, the 12 teams of League B occupy positions 13--24 according to the ranking in League B, the 15 teams of League C occupy positions 25--39 according to the ranking in League C, and the 16 teams of League D occupy positions 40--55 according to the ranking in League D.

The next stage is called the \href{https://en.wikipedia.org/wiki/UEFA_Euro_2020_qualifying}{UEFA Euro 2020 qualifiers}. The 55 teams are divided into five groups of five (Groups A--E) and five groups of six teams (Groups F--J), and they are seeded according to the overall UEFA Nations League ranking: teams 1--4 (from League A) are placed into the UNL Pot, teams 5--10 (from League A) are placed into Pot 1, teams 11--20 (from Leagues A and B) are placed into Pot 2, teams 21--30 (from Leagues B and C) are placed into Pot 3, teams 31--40 (from Leagues C and D) are placed into Pot 4, teams 41--50 (from League D) are placed into Pot 5, and teams 51--55 (from League D) are placed into Pot 6.
Then Groups A--D get one team from the UNL Pot each, and one team from Pots 2--5 each, Group E gets one team from Pots 1--5 each, while Groups F--J get one team from Pots 1--6 each. Thus the draw applies a standard procedure but the best four teams are guaranteed to be in the smaller groups of five.
There are also specific restrictions due to host nations, prohibited team clashes (because of political reasons), winter venue, and excessive travel \citep{UEFA2018d}. 
The groups are organized in a home-away (double) round-robin scheme and the first two teams qualify for the UEFA Euro 2020.

\begin{table}[t]
  \centering
  \caption{An overview of the qualification for the UEFA Euro 2020}
  \label{Table1}
  \rowcolors{3}{gray!20}{}
\begin{threeparttable}
    \begin{tabularx}{\textwidth}{CC >{\centering\arraybackslash}p{2.2cm} Cc} \toprule \hiderowcolors
    Rank from UEFA national team coefficient & League in the UEFA Nations League & UEFA Nations League overall rank & Place in the seeding of the UEFA Euro 2020 qualifiers & Remark \\ \midrule \showrowcolors
    1--12 & A     & 1--4 (GW) & UNL Pot & \begin{tabular}{@{}c@{}} drawn into a group of five \\ assured of at least play-offs \end{tabular} \\
    1--12  & A     & 5--10 & Pot 1 & --- \\
    1--12  & A     & 11--12 & Pot 2 & --- \\ \hline
    13--24 & B     & 13--16 (GW) & Pot 2 & assured of at least play-offs \\
    13--24 & B     & 17--20 & Pot 2 & --- \\
    13--24 & B     & 21--24 & Pot 3 & --- \\ \hline
    25--39 & C     & 25--28 (GW) & Pot 3 & assured of at least play-offs \\
    25--39 & C     & 29--30 & Pot 3 & --- \\
    25--39 & C     & 31--39 & Pot 4 & --- \\ \hline
    40--55 & D     & 40 (GW) & Pot 4 & assured of at least play-offs \\
    40--55 & D     & 41--43 (GW) & Pot 5 & assured of at least play-offs \\
    40--55 & D     & 44--50 & Pot 5 & --- \\
    40--55 & D     & 51--55 & Pot 6 & --- \\ \bottomrule
    \end{tabularx}
\begin{tablenotes} \footnotesize
\item GW stands for group winner in the UEFA Nations League
\end{tablenotes}
\end{threeparttable}
\end{table}

Table~\ref{Table1} provides a short overview of the 2018/19 UEFA Nations League and the UEFA Euro 2020 qualifiers.

The four remaining berths are filled through the \href{https://en.wikipedia.org/wiki/UEFA_Euro_2020_qualifying_play-offs}{UEFA Euro 2020 qualifying play-offs}.
Contrary to the previous UEFA European Championships, teams do not advance to the play-offs from the qualifying group stage, but they are selected on the basis of the overall UEFA Nations League ranking according to the \emph{team selection} rule \citep[Article~16.02]{UEFA2018c}: the four group winners of each league enter the play-offs unless they have already qualified when they are substituted by the next best-ranked team in the relevant league ranking that is available for the play-offs.
The 16 teams are divided into four paths containing four teams each according to the \emph{path formation} rule \citep[Article~16.03]{UEFA2018c}. This requires that no group winners face any team from a higher-ranked league, and leagues with at least four teams in the play-offs form their own path with four teams from the league.
In any play-off path, the highest-ranked team is matched with the lowest-ranked team and the two middle teams are matched against each other in the semifinals, where the higher-ranked teams play at home. The final is contested by the winners of the semifinals and is hosted by the winner of a semifinal drawn in advance. The winner of the final advances to the UEFA Euro 2020.

Consequently, at least one team from each league participates in the final tournament: if no team qualifies through the qualifying group stage, then the group winners of the league form an own path. This is especially important for League D where teams have a low chance to advance to the UEFA Euro 2020 through the qualifying group stage. For example, the four group winners in League D of the 2018/19 UEFA Nations League (Georgia, North Macedonia, Kosovo, Belarus) contest the slot available for League D. 
The whole procedure is explained by three possible scenarios in \citet{UEFA2017g}.
Nonetheless, the rules are contradictory and may lead to an unfair formulation of play-off paths \citep{Csato2020g}.

To summarise, the qualification process consists of the following stages:
\begin{enumerate}
\item
The 55 teams are allocated into Leagues A--D on the basis of the ranking from the initial UEFA national team coefficients.
\item
Four groups in each league are drawn on the basis of the ranking from the initial UEFA national team coefficients.
\item
Matches of the 2018/19 UEFA Nations League are played, and the results determine the overall UEFA Nations League ranking.
\item
Groups of the UEFA Euro 2020 qualifiers are drawn on the basis of the overall UEFA Nations League ranking.
\item
Matches of the UEFA Euro 2020 qualifiers are played, the first two teams from each group qualify (altogether 20 teams enter the UEFA Euro 2020).
\item
16 teams that failed to qualify through the UEFA Euro 2020 qualifiers are selected on the basis of the overall UEFA Nations League ranking by the team selection rule.
\item \label{Path_formation}
The 16 contestants of the UEFA Euro 2020 qualifying play-offs are divided into four paths of four teams each according to the path formation rule.
\item
Matches of the UEFA Euro 2020 qualifying play-offs are played, the winner of each path qualify (further four teams enter the UEFA Euro 2020).
\end{enumerate}

\section{Methodology} \label{Sec3}

We attempt to quantify the probability of qualification for the UEFA Euro 2020 for each UEFA member states, which immediately provides the tournament metric to be analyzed.
Our computer code closely follows the relevant UEFA regulations \citep{UEFA2018c, UEFA2018e} with some minor differences:
\begin{itemize}
\item
The first four places 1--4 of the overall UEFA Nations League ranking are determined on the basis of UEFA Nations League group results, and not through the \href{https://en.wikipedia.org/wiki/2019_UEFA_Nations_League_Finals}{(2019) UEFA Nations League Finals}.
The draw of the groups in the UEFA Euro 2020 qualifiers shows that this does not affect the outcome of the simulation.
\item
The specific restrictions in the draw of the UEFA Euro 2020 qualifying group stage \citep{UEFA2018d} are ignored because considering all of them would be cumbersome and supposedly would influence the results only marginally.
\item
The theoretical contradiction in the rules of the play-offs is avoided through the proposal of \citet[Section~4.1]{Csato2020g}.
This has essentially no effect on the tournament metric because at least 13 teams should qualify from the lowest-ranked associations of League D, which is highly improbable.
\end{itemize}

Match outcomes are modeled such that the probability with which a given team defeats its opponent is fixed \emph{a priori}. That simple approach is chosen because:
(1) the national teams play few matches in a year; 
(2) the methodology does not cover our main message; and
(3) the qualitative findings are so robust that they should be reliably reproduced by any reasonable prediction model.

The fixed winning probabilities are derived from the \href{https://en.wikipedia.org/wiki/World_Football_Elo_Ratings}{World Football Elo Ratings}, published regularly on the website \href{http://eloratings.net/}{eloratings.net}. 
Although there exists no single nor any official Elo ranking for football teams, Elo-inspired ratings seem to have the highest prediction power \citep{LasekSzlavikBhulai2013}. The overhauled formula for the FIFA/Coca-Cola World Ranking, introduced in June 2018, also uses the Elo method of calculation.
Elo rating immediately provides win expectancy $W_e$ according to the formula:
\begin{equation} \label{eq1}
W_e = \frac{1}{1 + 10^{-d/s}},
\end{equation}
where $d$ is the difference in the Elo ratings of the two teams, increased by 100 points for a team playing at home, while $s = 400$ is a scaling parameter.
Both the value of the home advantage and the scaling parameter $s$ comes from the system of World Football Elo Ratings (see \href{http://eloratings.net/about}{http://eloratings.net/about}).

For a match between teams $i$ and $j$, the win probability $w_i$ of team $i$ is computed by formula \eqref{eq1}, and a random number $r$ between 0 and 1 is generated. If $r < w_i$, then $i$ wins, otherwise, $j$ wins. Note that $w_i + w_j = 1$.
Draws (ties) in the matches are prohibited as further arbitrary assumptions would be necessary to obtain the probability of that outcome from formula \eqref{eq1}.
Ties in the group and league rankings are broken randomly.

Fixing home advantage at 100 points has a limited influence on the outcome of the simulations because both the 2018/19 UEFA Nations League and the UEFA Euro 2020 qualifiers are played in double (home-away) round-robin groups. Essentially, only the semifinals in the UEFA Euro 2020 qualifying play-off paths are affected by this choice, where the two higher-ranked teams are guaranteed to play at home.

\begin{sidewaystable}
\centering
\caption{The characteristics of national teams at the beginning of the qualification for the UEFA Euro 2020}
\label{Table2}
\rowcolors{4}{gray!20}{}
\begin{threeparttable}
    \begin{tabularx}{1\linewidth}{lccc C lccc} \toprule \hiderowcolors
    Team & UEFA rank & Elo rating & Elo rank &       & Team & UEFA rank & Elo rating & Elo rank \\ \midrule
    \multicolumn{4}{c}{\textbf{League A}} &       & \multicolumn{4}{c}{\textbf{League B}} \\ \bottomrule \showrowcolors
    Germany & 1     & 2109  & 1     &       & Austria & 13    & 1710  & 24 \\
    Portugal & 2     & 1995  & 3     &       & Wales & 14    & 1763  & 16 \\
    Belgium & 3     & 1927  & 6     &       & Russia & 15    & 1697  & 25 \\
    Spain & 4     & 2031  & 2     &       & Slovakia & 16    & 1748  & 17 \\
    France & 5     & 1989  & 4     &       & Sweden & 17    & 1825  & 13 \\
    England & 6     & 1933  & 5     &       & Ukraine & 18    & 1737  & 18 \\
    Switzerland & 7     & 1866  & 9     &       & Ireland & 19    & 1732  & 19 \\
    Italy & 8     & 1906  & 7     &       & Bosnia and Herzegovina & 20    & 1723  & 20 \\
    Poland & 9     & 1842  & 11    &       & Northern Ireland & 21    & 1674  & 27 \\
    Iceland & 10    & 1811  & 14    &       & Denmark & 22    & 1842  & 12 \\
    Croatia & 11    & 1856  & 10    &       & Czech Republic & 23    & 1713  & 22 \\
    Netherlands & 12    & 1895  & 8     &       & Turkey & 24    & 1712  & 23 \\ \bottomrule
    \multicolumn{4}{c}{\textbf{League C}} &       & \multicolumn{4}{c}{\textbf{League D}} \\ \toprule
    Hungary & 25    & 1611  & 32    &       & Azerbaijan & 40    & 1400  & 44 \\
    Romania & 26    & 1688  & 26    &       & North Macedonia & 41    & 1520  & 37 \\
    Scotland & 27    & 1720  & 21    &       & Belarus & 42    & 1497  & 39 \\
    Slovenia & 28    & 1639  & 29    &       & Georgia & 43    & 1483  & 40 \\
    Greece & 29    & 1661  & 28    &       & Armenia & 44    & 1480  & 41 \\
    Serbia & 30    & 1769  & 15    &       & Latvia & 45    & 1362  & 46 \\
    Albania & 31    & 1609  & 33    &       & Faroe Islands & 46    & 1281  & 50 \\
    Norway & 32    & 1581  & 35    &       & Luxembourg & 47    & 1321  & 49 \\
    Montenegro & 33    & 1614  & 31    &       & Kazakhstan & 48    & 1340  & 47 \\
    Israel & 34    & 1534  & 36    &       & Moldova & 49    & 1332  & 48 \\
    Bulgaria & 35    & 1620  & 30    &       & Liechtenstein & 50    & 1150  & 52 \\
    Finland & 36    & 1595  & 34    &       & Malta & 51    & 1216  & 51 \\
    Cyprus & 37    & 1410  & 42    &       & Andorra & 52    & 1012  & 54 \\
    Estonia & 38    & 1507  & 38    &       & Kosovo & 53    & 1371  & 45 \\
    Lithuania & 39    & 1406  & 43    &       & San Marino & 54    & 852   & 55 \\
          &       &       &       &       &  Gibraltar & 55    & 1079  & 53 \\ \toprule
    \end{tabularx}
\begin{tablenotes} \footnotesize
\item Source of the Elo rating: \href{https://www.international-football.net/elo-ratings-table?year=2017\&month=12\&day=06\&confed=UEFA}{https://www.international-football.net/elo-ratings-table?year=2017\&month=12\&day=06\&confed=UEFA} (6 December 2017)
\end{tablenotes}
\end{threeparttable}
\end{sidewaystable}

We have used Elo ratings as of 6 December 2017 since the seeding pots of the 2018/19 UEFA Nations League were announced on 7 December 2017.
They are reported in Table~\ref{Table2} besides the ranking derived from the UEFA national team coefficients, which provides the seeding of the 2018/19 UEFA Nations League.

A simulation run consists of playing all matches in the three subtournaments of the qualification process for the UEFA Euro 2020: the \href{https://en.wikipedia.org/wiki/2018\%E2\%80\%9319_UEFA_Nations_League}{2018/19 UEFA Nations League}, the \href{https://en.wikipedia.org/wiki/UEFA_Euro_2020_qualifying}{UEFA Euro 2020 qualifiers}, and the \href{https://en.wikipedia.org/wiki/UEFA_Euro_2020_qualifying_play-offs}{UEFA Euro 2020 qualifying play-offs}. Since it is only a single realization of random variables, one million iterations have been considered to obtain exact expected values.

\section{Results} \label{Sec4}

\begin{figure}[t]
\centering

\begin{tikzpicture}
\begin{axis}[width = 1\textwidth, 
height = 0.6\textwidth,
xmin = 825,
xmax = 2175,
ymin = -0.02,
ymax = 1.02,
ymajorgrids,
/pgf/number format/.cd,1000 sep={},
xlabel = Elo rating,
xlabel style = {font =\small},
legend entries = {League A$\qquad$,League B$\qquad$,League C$\qquad$,League D},
legend style = {at = {(0.5,-0.15)},anchor = north,legend columns = 4,font = \small}
]
\addplot[black,thick,only marks,mark=otimes*, mark size=2pt] coordinates {
(2109,0.998222)
(1995,0.988677)
(1927,0.967953)
(2031,0.993423)
(1989,0.987279)
(1933,0.969781)
(1866,0.926155)
(1906,0.956355)
(1842,0.897824)
(1811,0.855216)
(1856,0.914073)
(1895,0.947248)
};

\addplot[ForestGreen,very thick,only marks,mark=x, mark size=3pt] coordinates {
(1710,0.564022)
(1763,0.705899)
(1697,0.527145)
(1748,0.667663)
(1825,0.837816)
(1737,0.643438)
(1732,0.629879)
(1723,0.60505)
(1674,0.461385)
(1842,0.864133)
(1713,0.573332)
(1712,0.569296)
};

\addplot[red,thick,only marks,mark=diamond*, mark size=3pt] coordinates {
(1611,0.234095)
(1688,0.45028)
(1720,0.547951)
(1639,0.30631)
(1661,0.366152)
(1769,0.687713)
(1609,0.227118)
(1581,0.168441)
(1614,0.224132)
(1534,0.090503)
(1620,0.238851)
(1595,0.184324)
(1410,0.013527)
(1507,0.05829)
(1406,0.012605)
};

\addplot[blue,thick,only marks,mark=star, mark size=3pt] coordinates {
(1400,0.084713)
(1520,0.294993)
(1497,0.240655)
(1483,0.210226)
(1480,0.162817)
(1362,0.038263)
(1281,0.010959)
(1321,0.020911)
(1340,0.022303)
(1332,0.019541)
(1150,0.000621)
(1216,0.002508)
(1012,0.000009)
(1371,0.029862)
(1079,0.000063)
(852,0)
};
\end{axis}
\end{tikzpicture}

\caption[The probability of qualification for the UEFA Euro 2020]{The probability of qualification for the UEFA Euro 2020}
\label{Fig1}

\end{figure}
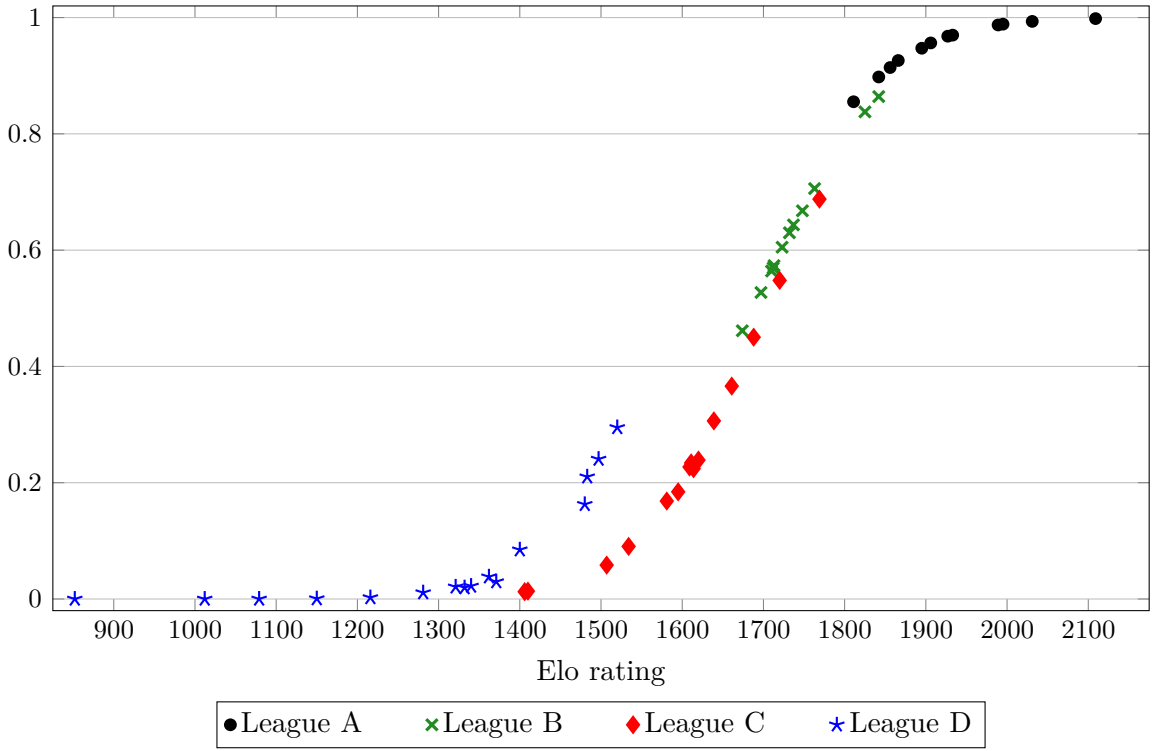

Figure~\ref{Fig1} plots the probability of qualification for the 55 teams as a function of their Elo rating. Intuitively, a team allocated into a weaker league, League B (C) instead of League A (B), has a smaller chance to qualify for the UEFA Euro 2020 \emph{ceteris paribus}. However, this is completely reversed in the comparison of Leagues C and D.

\begin{figure}[ht!]
\centering

\begin{subfigure}{\textwidth}
\centering
\caption{Probability of qualification through the UEFA Euro 2020 qualifiers}
\label{Fig2a}
\begin{tikzpicture}
\begin{axis}[width = 1\textwidth, 
height = 0.6\textwidth,
xmin = 825,
xmax = 2175,
ymin = -0.02,
ymax = 1.02,
ymajorgrids,
/pgf/number format/.cd,1000 sep={},
xlabel = Elo rating,
xlabel style = {font =\small},
legend entries = {League A$\qquad$,League B$\qquad$,League C$\qquad$,League D},
legend style = {at = {(0.5,-0.15)},anchor = north,legend columns = 4,font = \small}
]
\addplot[black,thick,only marks,mark=otimes*, mark size=2pt] coordinates {
(2109,0.992452)
(1995,0.968308)
(1927,0.929013)
(2031,0.979461)
(1989,0.965365)
(1933,0.932219)
(1866,0.863553)
(1906,0.909541)
(1842,0.823941)
(1811,0.767402)
(1856,0.845636)
(1895,0.894716)
};

\addplot[ForestGreen,very thick,only marks,mark=x, mark size=3pt] coordinates {
(1710,0.451593)
(1763,0.587779)
(1697,0.418411)
(1748,0.549133)
(1825,0.730743)
(1737,0.525202)
(1732,0.51245)
(1723,0.489207)
(1674,0.359712)
(1842,0.763424)
(1713,0.459471)
(1712,0.455696)
};

\addplot[red,thick,only marks,mark=diamond*, mark size=3pt] coordinates {
(1611,0.160028)
(1688,0.310004)
(1720,0.38423)
(1639,0.208825)
(1661,0.251379)
(1769,0.502829)
(1609,0.156297)
(1581,0.116179)
(1614,0.160673)
(1534,0.06531)
(1620,0.170749)
(1595,0.132515)
(1410,0.01145)
(1507,0.045969)
(1406,0.010691)
};

\addplot[blue,thick,only marks,mark=star, mark size=3pt] coordinates {
(1400,0.006496)
(1520,0.041687)
(1497,0.030183)
(1483,0.024433)
(1480,0.022949)
(1362,0.003114)
(1281,0.00065)
(1321,0.001429)
(1340,0.002109)
(1332,0.00175)
(1150,0.000027)
(1216,0.000142)
(1012,0.000002)
(1371,0.003469)
(1079,0.000004)
(852,0)
};
\end{axis}
\end{tikzpicture}
\end{subfigure}

\vspace{0.5cm}
\begin{subfigure}{\textwidth}
\centering
\caption{Probability of qualification through the UEFA Euro 2020 qualifying play-offs}
\label{Fig2b}
\begin{tikzpicture}
\begin{axis}[width = 1\textwidth, 
height = 0.6\textwidth,
xmin = 825,
xmax = 2175,
ymin = -0.01,
ymax = 0.31,
ymajorgrids,
yticklabel style = {/pgf/number format/fixed},
/pgf/number format/.cd,1000 sep={},
xlabel = Elo rating,
xlabel style = {font =\small},
legend entries = {League A$\qquad$,League B$\qquad$,League C$\qquad$,League D},
legend style = {at = {(0.5,-0.15)},anchor = north,legend columns = 4,font = \small}
]
\addplot[black,thick,only marks,mark=otimes*, mark size=2pt] coordinates {
(2109,0.00577)
(1995,0.020369)
(1927,0.03894)
(2031,0.013962)
(1989,0.021914)
(1933,0.037562)
(1866,0.062602)
(1906,0.046814)
(1842,0.073883)
(1811,0.087814)
(1856,0.068437)
(1895,0.052532)
};

\addplot[ForestGreen,very thick,only marks,mark=x, mark size=3pt] coordinates {
(1710,0.112429)
(1763,0.11812)
(1697,0.108734)
(1748,0.11853)
(1825,0.107073)
(1737,0.118236)
(1732,0.117429)
(1723,0.115843)
(1674,0.101673)
(1842,0.100709)
(1713,0.113861)
(1712,0.1136)
};

\addplot[red,thick,only marks,mark=diamond*, mark size=3pt] coordinates {
(1611,0.074067)
(1688,0.140276)
(1720,0.163721)
(1639,0.097485)
(1661,0.114773)
(1769,0.184884)
(1609,0.070821)
(1581,0.052262)
(1614,0.063459)
(1534,0.025193)
(1620,0.068102)
(1595,0.051809)
(1410,0.002077)
(1507,0.012321)
(1406,0.001914)
};

\addplot[blue,thick,only marks,mark=star, mark size=3pt] coordinates {
(1400,0.078217)
(1520,0.253306)
(1497,0.210472)
(1483,0.185793)
(1480,0.139868)
(1362,0.035149)
(1281,0.010309)
(1321,0.019482)
(1340,0.020194)
(1332,0.017791)
(1150,0.000594)
(1216,0.002366)
(1012,0.000007)
(1371,0.026393)
(1079,0.000059)
(852,0)
};
\end{axis}
\end{tikzpicture}
\end{subfigure}

\caption{The decomposed probability of qualification for the UEFA Euro 2020}
\label{Fig2}

\end{figure}

According to Figure~\ref{Fig2}, being in a lower-ranked league has two separate effects:
\begin{itemize}
\item
the probability of qualification through the qualifying group stage is decreased because being in a weaker league reduces (eliminates) the chance of obtaining a place in a stronger pot, hence the team should play against better teams on average in the UEFA Euro 2020 qualifiers (Figure~\ref{Fig2a}); but
\item
the probability of qualification through the play-offs is increased because the team can easier win its Nations League group, which guarantees a place in the play-off path of its league (Figure~\ref{Fig2b}).
\end{itemize}
When direct qualification through the UEFA Euro 2020 qualifiers has a low probability, that is, in the comparison of the bottom of League C with the top of League D for the given distribution of teams' strength, the second effect dominates the first, hence having a worse rank at the beginning of the 2018/19 UEFA Nations League remarkably raises the probability of qualification for the UEFA Euro 2020.

Figure~\ref{Fig2b} can be somewhat misleading: the probability of qualification through the play-offs is a non-increasing function of the Elo rating in Leagues A and B only because these teams usually qualify through the qualifying group stage and not through the play-offs.
Naturally, the \emph{conditional} probability of qualifying through the play-offs provided that a team participates in the play-offs depends monotonically on the Elo rating.

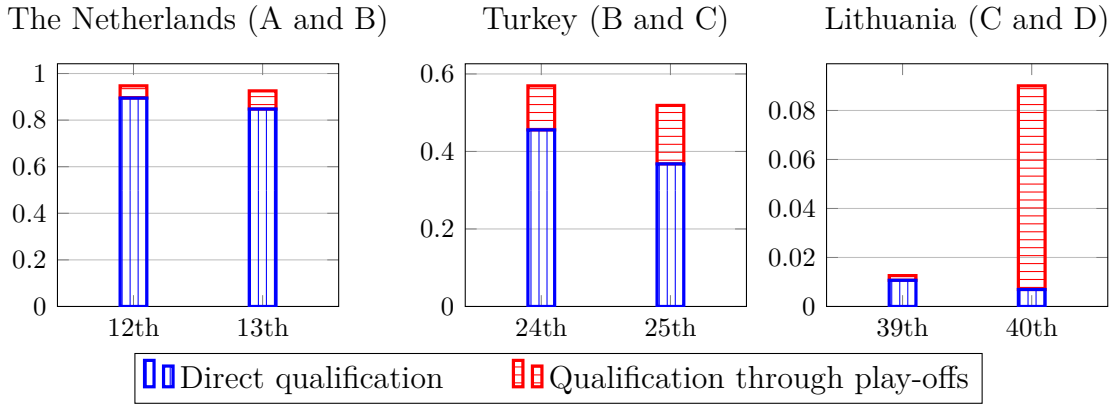
\begin{figure}[t]
\centering

\begin{tikzpicture}
\begin{axis}[width = 0.33\textwidth, 
height = 0.3\textwidth,
title = {The Netherlands (A and B)},
symbolic x coords = {12th, 13th},
xtick = data,
tick label style = {/pgf/number format/.cd, scaled ticks = false, fixed},
ybar stacked,
ymin = 0,
ymajorgrids = true,
enlarge x limits = {abs = 1cm},
]
\addplot [blue, pattern color = blue, pattern = vertical lines, very thick] coordinates{
(12th,0.894716)
(13th,0.847558)
};
\addplot [red, pattern color = red, pattern = horizontal lines, very thick] coordinates{
(12th,0.052532)
(13th,0.078014)
};
\end{axis}
\end{tikzpicture}
\begin{tikzpicture}
\begin{axis}[width = 0.33\textwidth, 
height = 0.3\textwidth,
title = Turkey (B and C),
symbolic x coords = {24th, 25th},
xtick = data,
tick label style = {/pgf/number format/.cd, scaled ticks = false, fixed},
ybar stacked,
ymin = 0,
ymajorgrids = true,
enlarge x limits = {abs = 1cm},
]
\addplot [blue, pattern color = blue, pattern = vertical lines, very thick] coordinates{
(24th,0.455696)
(25th,0.367929)
};
\addplot [red, pattern color = red, pattern = horizontal lines, very thick] coordinates{
(24th,0.1136)
(25th,0.150847)
};
\end{axis}
\end{tikzpicture}
\begin{tikzpicture}
\begin{axis}[width = 0.33\textwidth, 
height = 0.3\textwidth,
title = Lithuania (C and D),
symbolic x coords = {39th, 40th},
xtick = data,
tick label style = {/pgf/number format/.cd, scaled ticks = false, fixed},
ybar stacked,
ymin = 0,
ymajorgrids = true,
enlarge x limits = {abs = 1cm},
]
\addplot [blue, pattern color = blue, pattern = vertical lines, very thick] coordinates{
(39th,0.010691)
(40th,0.006931)
};
\addplot [red, pattern color = red, pattern = horizontal lines, very thick] coordinates{
(39th,0.001914)
(40th,0.083131)
};
\end{axis}
\end{tikzpicture}

\vspace{-0.4cm}
\begin{center}
\begin{tikzpicture}
	\begin{customlegend}[legend columns = 2, legend entries = {Direct qualification$\qquad$, Qualification through play-offs},
	legend image code/.code={\draw[fill] (0cm,-0.1cm) rectangle (0.125cm,0.25cm) (0.25cm,-0.1cm) rectangle (0.375cm,0.175cm);}]
        \addlegendimage{color = blue, pattern = vertical lines, pattern color = blue, very thick}
        \addlegendimage{color = red, pattern = horizontal lines, pattern color = red, very thick} 
	\end{customlegend}
\end{tikzpicture}
\end{center}
\vspace{-0.5cm}

\caption{The probability of qualification for teams at the boundary of two leagues}
\label{Fig3}
\end{figure}

To further illustrate the unfairness of the qualification, Figure~\ref{Fig3} presents the actual and hypothetical probabilities of qualification for three countries (be aware of the different individual scales on the y-axis, applied to increase visibility):
\begin{itemize}
\item
the Netherlands as the worst team in League A (12th) versus the best team in League B (13th);
\item
Turkey as the worst team in League B (24th) versus the best team in League C (25th);
\item
Lithuania as the worst team in League C (39th) versus the best team in League D (40th).
\end{itemize}
Being in a lower-ranked league decreases the probability of qualification by about 2.3\% for the Netherlands and by about 8.9\% for Turkey, but the same metric increases by more than seven-fold for Lithuania. While the probability of direct qualification becomes lower and the probability of qualification through the play-offs becomes higher in all cases, the latter effect dominates the former in the comparison of Leagues C and D, which may create an avenue for tanking.

The sensitivity of the calculations can be checked by modifying the scaling parameter $s$ in expression~\eqref{eq1}. Its original value of $400$ may be judged excessive: for example, it implies that the best team, Germany defeats the 10th (Croatia) with a probability of 81.1\% at a neutral field, and wins against the 30th (Bulgaria) with a probability of 93.4\%.
Furthermore, it is worth comparing the worst place of League C (39th) to the fifth place of League D (44th) as obtaining the 40th position by a strategic manipulation of the UEFA national team coefficient cannot be guaranteed. Note that there is no difference in the probability of qualification between the 37--39 positions, as well as between the 40--43 positions due to the seeding procedure in the UEFA Nations League.

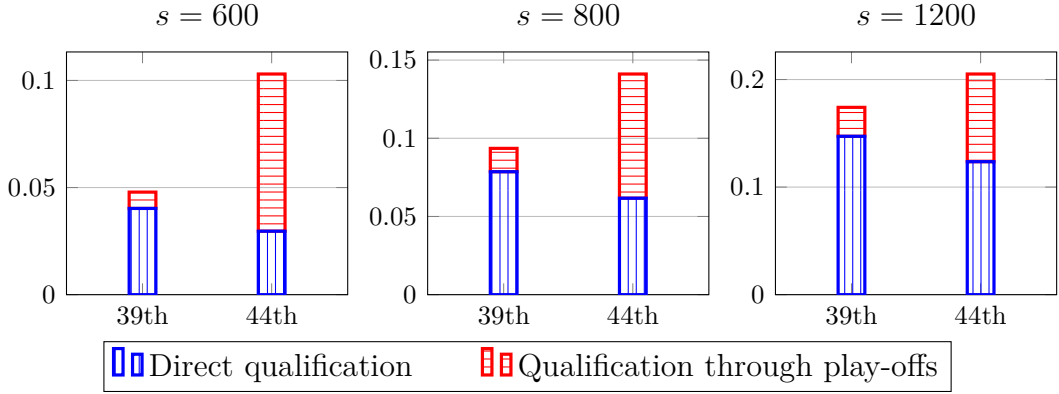
\begin{figure}[t]
\centering

\begin{tikzpicture}
\begin{axis}[width = 0.33\textwidth, 
height = 0.3\textwidth,
title = {$s = 600$},
symbolic x coords = {39th, 44th},
xtick = data,
tick label style = {/pgf/number format/.cd, scaled ticks = false, fixed},
ybar stacked,
ymin = 0,
ymajorgrids = true,
enlarge x limits = {abs = 1cm},
]
\addplot [blue, pattern color = blue, pattern = vertical lines, very thick] coordinates{
(39th,0.040255)
(44th,0.029632)
};
\addplot [red, pattern color = red, pattern = horizontal lines, very thick] coordinates{
(39th,0.007617)
(44th,0.073372)
};
\end{axis}
\end{tikzpicture}
\begin{tikzpicture}
\begin{axis}[width = 0.33\textwidth, 
height = 0.3\textwidth,
title = {$s = 800$},
symbolic x coords = {39th, 44th},
xtick = data,
tick label style = {/pgf/number format/.cd, scaled ticks = false, fixed},
ybar stacked,
ymin = 0,
ymajorgrids = true,
enlarge x limits = {abs = 1cm},
]
\addplot [blue, pattern color = blue, pattern = vertical lines, very thick] coordinates{
(39th,0.078616)
(44th,0.06169)
};
\addplot [red, pattern color = red, pattern = horizontal lines, very thick] coordinates{
(39th,0.014929)
(44th,0.079405)
};
\end{axis}
\end{tikzpicture}
\begin{tikzpicture}
\begin{axis}[width = 0.33\textwidth, 
height = 0.3\textwidth,
title = {$s = 1200$},
symbolic x coords = {39th, 44th},
xtick = data,
tick label style = {/pgf/number format/.cd, scaled ticks = false, fixed},
ybar stacked,
ymin = 0,
ymajorgrids = true,
enlarge x limits = {abs = 1cm},
]
\addplot [blue, pattern color = blue, pattern = vertical lines, very thick] coordinates{
(39th,0.14722)
(44th,0.123635)
};
\addplot [red, pattern color = red, pattern = horizontal lines, very thick] coordinates{
(39th,0.02691)
(44th,0.081483)
};
\end{axis}
\end{tikzpicture}

\vspace{-0.4cm}
\begin{center}
\begin{tikzpicture}
	\begin{customlegend}[legend columns = 2, legend entries = {Direct qualification$\qquad$, Qualification through play-offs},
	legend image code/.code={\draw[fill] (0cm,-0.1cm) rectangle (0.125cm,0.25cm) (0.25cm,-0.1cm) rectangle (0.375cm,0.175cm);}]
        \addlegendimage{color = blue, pattern = vertical lines, pattern color = blue, very thick}
        \addlegendimage{color = red, pattern = horizontal lines, pattern color = red, very thick} 
	\end{customlegend}
\end{tikzpicture}
\end{center}
\vspace{-0.5cm}

\caption{The probability of qualification for Lithuania: sensitivity analysis}
\label{Fig4}

\end{figure}

Thus three more competitive distributions of teams' strength have been considered and plotted in Figure~\ref{Fig4} (again, be aware of the different individual scales on the y-axis). Although the advantage of the 44th over the 39th place decreases when inequality is reduced, it remains substantial. Therefore any team has a strong incentive to avoid being the 37th, the 38th, and the 39th in the UEFA ranking underlying the construction of the 2018/19 UEFA Nations League.


\section{How to achieve fairness} \label{Sec5}

We have devised two straightforward proposals to mitigate unfairness that still award group winners in the UEFA Nations League. For the sake of simplicity, only the last step of the qualification, the path formation of the UEFA Euro 2020 qualifying play-offs is modified (see item~\ref{Path_formation} at the end of Section~\ref{Sec2}).
In particular, two alternative policies to the regular path formation rule are considered:
\begin{itemize}
\item
\emph{Random path formation}: the 16 teams of the play-offs are divided randomly into four paths.
\item
\emph{Seeded path formation}: the 16 teams of the play-offs are divided into four paths on the basis of the overall UEFA Nations League ranking under the traditional seeding regime, that is, the four highest-ranked teams are drawn randomly from Pot 1, the next four from Pot 2, the next four from Pot 3, and the four lowest-ranked from Pot 4.
\end{itemize}
While the four teams (usually the group winners) of League D always form a separate path under the regular path formation rule, they can play against higher-ranked teams if random path formation is applied (however, they still have some chance to avoid the strongest teams), and they should play against one of the four highest-ranked teams in the semifinal if seeded path formation is followed.
The modification of the path formation rule affects only the probability of qualification through the play-offs (i.e.\ Figure~\ref{Fig2b}) but not the probability of qualification through the qualifying group stage (plotted in Figure~\ref{Fig2a}).

\begin{figure}[ht!]
\centering

\begin{subfigure}{\textwidth}
\centering
\caption{Random path formation}
\label{Fig5a}
\begin{tikzpicture}
\begin{axis}[width = 1\textwidth, 
height = 0.6\textwidth,
xmin = 825,
xmax = 2175,
ymin = -0.02,
ymax = 1.02,
ymajorgrids,
/pgf/number format/.cd,1000 sep={},
xlabel = Elo rating,
xlabel style = {font =\small},
legend entries = {League A$\qquad$,League B$\qquad$,League C$\qquad$,League D},
legend style = {at = {(0.5,-0.15)},anchor = north,legend columns = 4,font = \small}
]
\addplot[black,thick,only marks,mark=otimes*, mark size=2pt] coordinates {
(2109,0.998926)
(1995,0.992506)
(1927,0.977925)
(2031,0.995837)
(1989,0.991526)
(1933,0.979339)
(1866,0.947201)
(1906,0.969637)
(1842,0.926151)
(1811,0.892346)
(1856,0.937816)
(1895,0.963097)
};

\addplot[ForestGreen,very thick,only marks,mark=x, mark size=3pt] coordinates {
(1710,0.641225)
(1763,0.763922)
(1697,0.60839)
(1748,0.73094)
(1825,0.871501)
(1737,0.709798)
(1732,0.698663)
(1723,0.676641)
(1674,0.548682)
(1842,0.8929)
(1713,0.649203)
(1712,0.645612)
};

\addplot[red,thick,only marks,mark=diamond*, mark size=3pt] coordinates {
(1611,0.23381)
(1688,0.430972)
(1720,0.520648)
(1639,0.299338)
(1661,0.355823)
(1769,0.653033)
(1609,0.226815)
(1581,0.17009)
(1614,0.223473)
(1534,0.091976)
(1620,0.237944)
(1595,0.185178)
(1410,0.013763)
(1507,0.058882)
(1406,0.01287)
};

\addplot[blue,thick,only marks,mark=star, mark size=3pt] coordinates {
(1400,0.015896)
(1520,0.080065)
(1497,0.060523)
(1483,0.050432)
(1480,0.042018)
(1362,0.006799)
(1281,0.001606)
(1321,0.003381)
(1340,0.004277)
(1332,0.003598)
(1150,0.000073)
(1216,0.000348)
(1012,0.000002)
(1371,0.006576)
(1079,0.000007)
(852,0)
};
\end{axis}
\end{tikzpicture}
\end{subfigure}

\vspace{0.5cm}
\begin{subfigure}{\textwidth}
\centering
\caption{Seeded path formation}
\label{Fig5b}
\begin{tikzpicture}
\begin{axis}[width = 1\textwidth, 
height = 0.6\textwidth,
xmin = 825,
xmax = 2175,
ymin = -0.02,
ymax = 1.02,
ymajorgrids,
yticklabel style = {/pgf/number format/fixed},
/pgf/number format/.cd,1000 sep={},
xlabel = Elo rating,
xlabel style = {font =\small},
legend entries = {League A$\qquad$,League B$\qquad$,League C$\qquad$,League D},
legend style = {at = {(0.5,-0.15)},anchor = north,legend columns = 4,font = \small}
]
\addplot[black,thick,only marks,mark=otimes*, mark size=2pt] coordinates {
(2109,0.99918)
(1995,0.993904)
(1927,0.981893)
(2031,0.996749)
(1989,0.993202)
(1933,0.983031)
(1866,0.955305)
(1906,0.974841)
(1842,0.937358)
(1811,0.907393)
(1856,0.947862)
(1895,0.969337)
};

\addplot[ForestGreen,very thick,only marks,mark=x, mark size=3pt] coordinates {
(1710,0.662847)
(1763,0.783172)
(1697,0.630505)
(1748,0.751017)
(1825,0.885997)
(1737,0.730815)
(1732,0.719575)
(1723,0.698868)
(1674,0.568805)
(1842,0.90535)
(1713,0.670693)
(1712,0.667842)
};

\addplot[red,thick,only marks,mark=diamond*, mark size=3pt] coordinates {
(1611,0.209733)
(1688,0.408488)
(1720,0.502458)
(1639,0.275073)
(1661,0.330368)
(1769,0.639722)
(1609,0.20322)
(1581,0.150717)
(1614,0.203188)
(1534,0.080804)
(1620,0.216036)
(1595,0.166396)
(1410,0.012541)
(1507,0.0532)
(1406,0.011725)
};

\addplot[blue,thick,only marks,mark=star, mark size=3pt] coordinates {
(1400,0.011904)
(1520,0.065421)
(1497,0.04865)
(1483,0.040132)
(1480,0.034489)
(1362,0.005115)
(1281,0.001152)
(1321,0.002408)
(1340,0.003247)
(1332,0.002728)
(1150,0.000047)
(1216,0.000226)
(1012,0.000002)
(1371,0.005262)
(1079,0.000007)
(852,0)
};
\end{axis}
\end{tikzpicture}
\end{subfigure}

\caption{The probability of qualification: alternative path formations}
\label{Fig5}

\end{figure}

Figure~\ref{Fig5} shows the overall probability of qualification for the UEFA Euro 2020 under the proposed random (Figure~\ref{Fig5a}) and seeded (Figure~\ref{Fig5b}) path formation regimes. It remains better to be in League A (B) than in League B (C) and the difference between these leagues increases as path formation moves farther from its original concept.
In addition, unfairness becomes less serious in the case of random path formation and can be almost wiped out with seeded path formation.

\begin{figure}[t]
\centering

\begin{tikzpicture}
\begin{axis}[width = 0.33\textwidth, 
height = 0.3\textwidth,
title = Regular path formation,
symbolic x coords = {39th, 40th},
xtick = data,
tick label style = {/pgf/number format/.cd, scaled ticks = false, fixed},
ybar stacked,
ymin = 0,
ymajorgrids = true,
enlarge x limits = {abs = 1cm},
]
\addplot [blue, pattern color = blue, pattern = vertical lines, very thick] coordinates{
(39th,0.010691)
(40th,0.006931)
};
\addplot [red, pattern color = red, pattern = horizontal lines, very thick] coordinates{
(39th,0.001914)
(40th,0.083131)
};
\end{axis}
\end{tikzpicture}
\begin{tikzpicture}
\begin{axis}[width = 0.33\textwidth, 
height = 0.3\textwidth,
title = Random path formation,
symbolic x coords = {39th, 40th},
xtick = data,
tick label style = {/pgf/number format/.cd, scaled ticks = false, fixed},
ybar stacked,
ymin = 0,
ymajorgrids = true,
enlarge x limits = {abs = 1cm},
]
\addplot [blue, pattern color = blue, pattern = vertical lines, very thick] coordinates{
(39th,0.010691)
(40th,0.006931)
};
\addplot [red, pattern color = red, pattern = horizontal lines, very thick] coordinates{
(39th,0.002179)
(40th,0.01029)
};
\end{axis}
\end{tikzpicture}
\begin{tikzpicture}
\begin{axis}[width = 0.33\textwidth, 
height = 0.3\textwidth,
title = Seeded path formation,
symbolic x coords = {39th, 40th},
xtick = data,
tick label style = {/pgf/number format/.cd, scaled ticks = false, fixed, precision = 3},
ybar stacked,
ymin = 0,
ymajorgrids = true,
enlarge x limits = {abs = 1cm},
]
\addplot [blue, pattern color = blue, pattern = vertical lines, very thick] coordinates{
(39th,0.010691)
(40th,0.006931)
};
\addplot [red, pattern color = red, pattern = horizontal lines, very thick] coordinates{
(39th,0.001034)
(40th,0.005807)
};
\end{axis}
\end{tikzpicture}

\vspace{-0.4cm}
\begin{center}
\begin{tikzpicture}
	\begin{customlegend}[legend columns = 2, legend entries = {Direct qualification$\qquad$, Qualification through play-offs},
	legend image code/.code={\draw[fill] (0cm,-0.1cm) rectangle (0.125cm,0.25cm) (0.25cm,-0.1cm) rectangle (0.375cm,0.175cm);}]
        \addlegendimage{color = blue, pattern = vertical lines, pattern color = blue, very thick}
        \addlegendimage{color = red, pattern = horizontal lines, pattern color = red, very thick} 
	\end{customlegend}
\end{tikzpicture}
\end{center}
\vspace{-0.5cm}

\caption{The probability of qualification for Lithuania: various path formations}
\label{Fig6}

\end{figure}
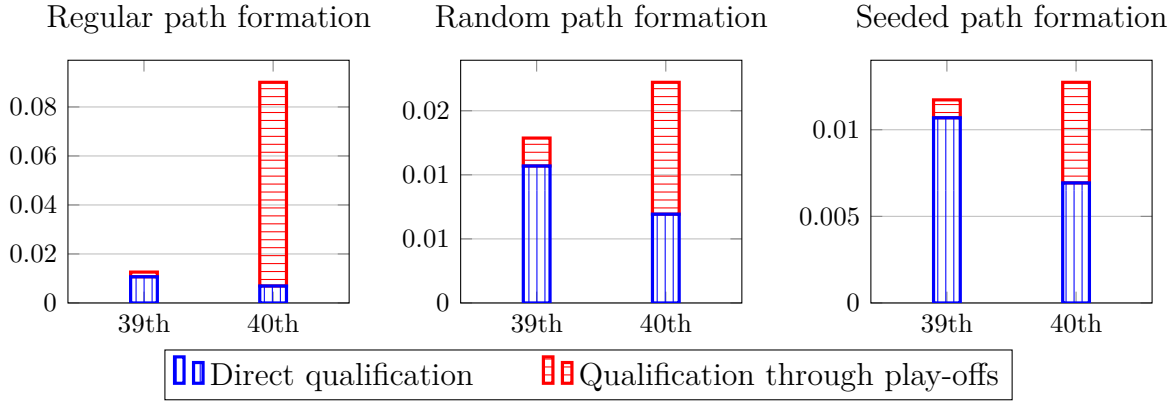

This implication is reinforced by Figure~\ref{Fig6}, which presents the probability of qualification for Lithuania, a team at the boundary of Leagues C and D. Since our proposals do not affect the qualifying group stage, the height of the blue columns with vertical lines are the same but the scale on the vertical axis is changed in order to highlight the differences. Playing in League D (40th position) increases the probability of qualification by a factor of $7.14$, $1.34$, and $1.09$ under the three path formation rules called regular, random, and seeded, respectively.

To summarize, the unfairness of the tournament design can be remarkably reduced or even eliminated with a slight policy change in the path formation of the UEFA Euro 2020 qualifying play-offs. Since the modifications do not influence the team selection rule, all contestants remain interested in performing well in the UEFA Nations League.

\section{Theoretical confirmation} \label{Sec6}

One can still say that the above finding of unfairness is mostly driven by the dataset (the actual distribution of teams' strength) and not by the competition format itself.
Therefore, another kind of sensitivity analysis is provided on the basis of win expectancies given by formula~\eqref{eq1}. In particular, three models called Difference 0, 10, and 20, respectively, are investigated: the Elo rating of the $28$th ranked middle team is fixed at $1500$, while team $i$ has an Elo rating of $1500 + (28-i) \Delta$, where $\Delta \in \{ 0, 10, 20 \}$ corresponds to the name of the probabilistic model.

\begin{figure}[t]
\centering

\begin{tikzpicture}
\begin{axis}[width = 1\textwidth, 
height = 0.6\textwidth,
xmin = 0,
xmax = 56,
ymin = -0.02,
ymax = 1.02,
ymajorgrids,
xlabel = Team rank,
xlabel style = {font =\small},
legend style = {at = {(0.5,-0.15)},anchor = north,legend columns = 4,font = \small}
]
\addlegendimage{black,thick,only marks,mark=otimes*, mark size=2pt}
\addlegendentry{League A$\qquad$}
\addplot[red,thick,only marks,mark=otimes*, mark size=2pt, forget plot] coordinates {
(1,0.450331)
(2,0.450211)
(3,0.450114)
(4,0.449991)
(5,0.449372)
(6,0.4503)
(7,0.450526)
(8,0.450265)
(9,0.449683)
(10,0.449585)
(11,0.450689)
(12,0.449863)
};

\addplot[blue,thick,only marks,mark=otimes*, mark size=2pt, forget plot] coordinates {
(1,0.969818)
(2,0.965165)
(3,0.959238)
(4,0.953304)
(5,0.946233)
(6,0.938056)
(7,0.929035)
(8,0.919853)
(9,0.907625)
(10,0.895556)
(11,0.88145)
(12,0.86772)
};

\addplot[ForestGreen,thick,only marks,mark=otimes*, mark size=2pt, forget plot] coordinates {
(1,0.998848)
(2,0.99841)
(3,0.99777)
(4,0.997041)
(5,0.99591)
(6,0.994369)
(7,0.992447)
(8,0.989992)
(9,0.98592)
(10,0.981462)
(11,0.975683)
(12,0.968446)
};

\addlegendimage{black,very thick,only marks,mark=x, mark size=3pt}
\addlegendentry{League B$\qquad$}
\addplot[red,very thick,only marks,mark=x, mark size=3pt, forget plot] coordinates {
(13,0.450188)
(14,0.449833)
(15,0.449728)
(16,0.450636)
(17,0.449833)
(18,0.449809)
(19,0.450258)
(20,0.449628)
(21,0.450202)
(22,0.450294)
(23,0.449935)
(24,0.448735)
};

\addplot[blue,very thick,only marks,mark=x, mark size=3pt, forget plot] coordinates {
(13,0.814502)
(14,0.793572)
(15,0.770402)
(16,0.747688)
(17,0.717412)
(18,0.691041)
(19,0.66302)
(20,0.633909)
(21,0.595206)
(22,0.566167)
(23,0.533034)
(24,0.5031)
};

\addplot[ForestGreen,very thick,only marks,mark=x, mark size=3pt, forget plot] coordinates {
(13,0.922648)
(14,0.903482)
(15,0.879661)
(16,0.853309)
(17,0.81402)
(18,0.776056)
(19,0.734513)
(20,0.68932)
(21,0.608409)
(22,0.555788)
(23,0.501273)
(24,0.448557)
};

\addlegendimage{black,thick,only marks,mark=diamond*, mark size=3pt}
\addlegendentry{League C$\qquad$}
\addplot[red,thick,only marks,mark=diamond*, mark size=3pt, forget plot] coordinates {
(25,0.433848)
(26,0.433525)
(27,0.434296)
(28,0.434285)
(29,0.433621)
(30,0.43354)
(31,0.433726)
(32,0.434687)
(33,0.434064)
(34,0.434464)
(35,0.433753)
(36,0.432889)
(37,0.430477)
(38,0.430454)
(39,0.431553)
};

\addplot[blue,thick,only marks,mark=diamond*, mark size=3pt, forget plot] coordinates {
(25,0.437257)
(26,0.404658)
(27,0.37349)
(28,0.341693)
(29,0.30338)
(30,0.275601)
(31,0.248266)
(32,0.225035)
(33,0.193962)
(34,0.173786)
(35,0.154426)
(36,0.137522)
(37,0.115836)
(38,0.101184)
(39,0.088396)
};

\addplot[ForestGreen,thick,only marks,mark=diamond*, mark size=3pt, forget plot] coordinates {
(25,0.448669)
(26,0.392292)
(27,0.337373)
(28,0.285773)
(29,0.226108)
(30,0.185418)
(31,0.150261)
(32,0.119457)
(33,0.08029)
(34,0.062353)
(35,0.047326)
(36,0.036133)
(37,0.023891)
(38,0.017692)
(39,0.012504)
};

\addlegendimage{black,thick,only marks,mark=star, mark size=3pt}
\addlegendentry{League D}
\addplot[red,thick,only marks,mark=star, mark size=3pt, forget plot] coordinates {
(40,0.418465)
(41,0.418877)
(42,0.418566)
(43,0.418127)
(44,0.419728)
(45,0.417922)
(46,0.419214)
(47,0.419511)
(48,0.418628)
(49,0.419807)
(50,0.418029)
(51,0.41893)
(52,0.418004)
(53,0.418839)
(54,0.418919)
(55,0.419243)
};

\addplot[blue,thick,only marks,mark=star, mark size=3pt, forget plot] coordinates {
(40,0.191703)
(41,0.170415)
(42,0.150947)
(43,0.132626)
(44,0.108007)
(45,0.094172)
(46,0.081951)
(47,0.071119)
(48,0.05579)
(49,0.047931)
(50,0.040809)
(51,0.034478)
(52,0.026511)
(53,0.022289)
(54,0.019015)
(55,0.015639)
};

\addplot[ForestGreen,thick,only marks,mark=star, mark size=3pt, forget plot] coordinates {
(40,0.240321)
(41,0.196376)
(42,0.158967)
(43,0.126508)
(44,0.081127)
(45,0.062203)
(46,0.047176)
(47,0.035533)
(48,0.020979)
(49,0.014821)
(50,0.010506)
(51,0.00737)
(52,0.003885)
(53,0.002549)
(54,0.001673)
(55,0.001132)
};
\draw (axis cs:12.5,\pgfkeysvalueof{/pgfplots/ymin})  -- (axis cs:12.5,\pgfkeysvalueof{/pgfplots/ymax});
\draw (axis cs:24.5,\pgfkeysvalueof{/pgfplots/ymin})  -- (axis cs:24.5,\pgfkeysvalueof{/pgfplots/ymax});
\draw (axis cs:39.5,\pgfkeysvalueof{/pgfplots/ymin})  -- (axis cs:39.5,\pgfkeysvalueof{/pgfplots/ymax});

\addlegendimage{line legend,very thick,red}
\addlegendentry{Difference 0$\qquad$}
\addlegendimage{line legend,very thick,blue}
\addlegendentry{Difference 10$\qquad$}
\addlegendimage{line legend,very thick,ForestGreen}
\addlegendentry{Difference 20$\qquad$}
\end{axis}
\end{tikzpicture}

\caption{Theoretical model: qualifying probability with regular path formation}
\label{Fig7}

\end{figure}
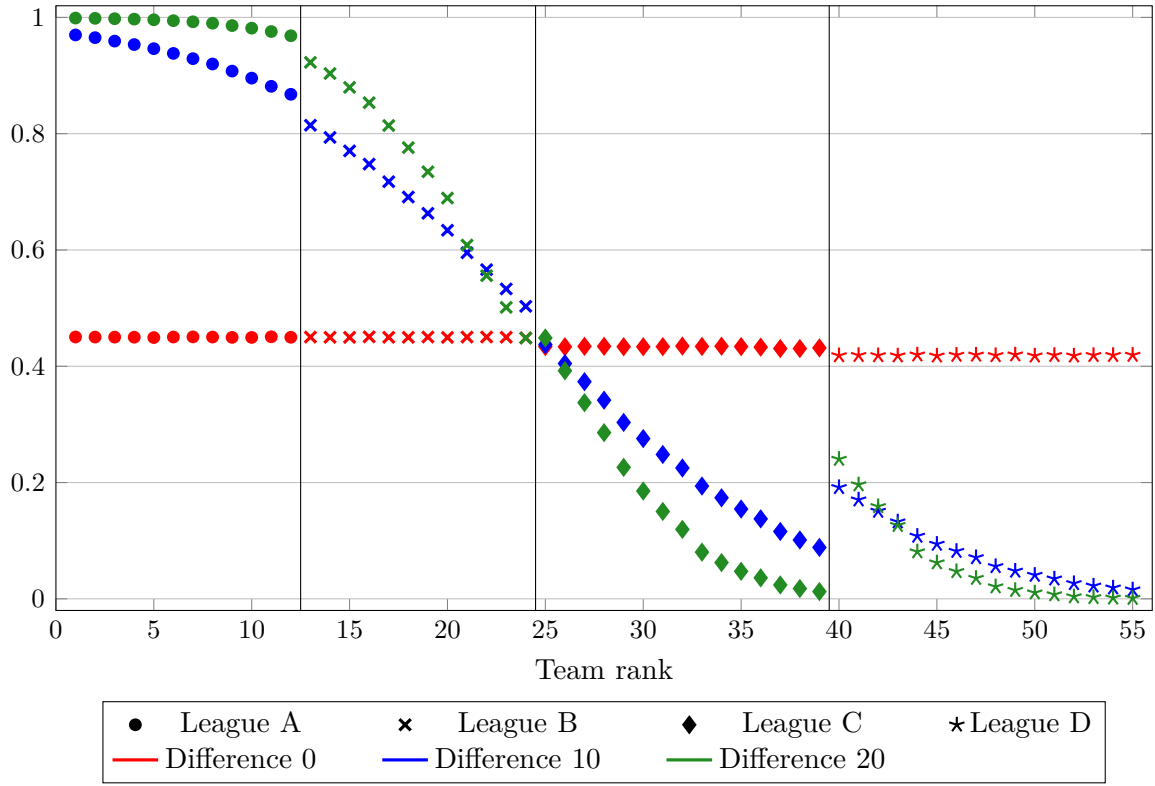

According to Figure~\ref{Fig7}, the probability of qualification is close to uniform around $24/55 \approx 0.4364$ if all teams are equally strong. Four factors imply differences among the teams:
\begin{itemize}
\item
the teams ranked $51$--$55$ are seeded in groups of six, hence a team from League D has to play against more teams on average in the UEFA Euro 2020 qualifiers;
\item
Leagues A and B have only 12 teams but League C consists of 15 and League D contains 16 teams, thus lower-ranked teams have a lower chance to win their Nations League group and advance to the qualifying play-offs;
\item
the team selection rule prefers higher-ranked teams in filling the vacant slots in the qualifying play-offs; and
\item
in each play-off path, the semifinals are hosted by the two higher-ranked teams.
\end{itemize}
In this case, the last two effects are marginal, that is, the teams of Leagues A and B have essentially the same probability to qualify.
In addition, the probability of direct qualification is the same for the teams of Leagues A, B, and C as only the first point affects the UEFA Euro 2020 qualifiers.

Contrarily, with increasing parameter $\Delta$, visible breaks appear at the boundary of leagues. The teams ranked $21$--$24$, the four weakest in League B, have a considerable disadvantage compared to the better teams of League B if $\Delta = 20$ because exactly 20 slots are allocated in the UEFA Euro 2020 qualifiers.
Unsurprisingly, participation in League A is favored to participation in League B, as well as League B is preferred to League C. On the other hand---analogously to the finding with the real data in Section~\ref{Sec4}---it is better to play in League D compared to League C. Unfairness grows in parallel with competitive imbalance. In the model of Difference 20, even the team at bottom of League B is somewhat worse off than the team at the top of League C, which implies that incentive incompatibility might be a problem at the boundary of the two middle leagues if the profile of teams' strength changes.

\begin{figure}[ht!]
\centering

\begin{subfigure}{\textwidth}
\centering
\caption{Random path formation}
\label{Fair_Euro2020_Fig14a}
\begin{tikzpicture}
\begin{axis}[width = 1\textwidth, 
height = 0.6\textwidth,
xmin = 0,
xmax = 56,
ymin = -0.02,
ymax = 1.02,
ymajorgrids,
xlabel = Team rank,
xlabel style = {font =\small},
legend style = {at = {(0.5,-0.15)},anchor = north,legend columns = 4,font = \small}
]
\addlegendimage{black,thick,only marks,mark=otimes*, mark size=2pt}
\addlegendentry{League A$\qquad$}
\addplot[red,thick,only marks,mark=otimes*, mark size=2pt, forget plot] coordinates {
(1,0.472594)
(2,0.471827)
(3,0.471994)
(4,0.471866)
(5,0.471119)
(6,0.472093)
(7,0.472777)
(8,0.472426)
(9,0.4713)
(10,0.471949)
(11,0.472449)
(12,0.471632)
};

\addplot[blue,thick,only marks,mark=otimes*, mark size=2pt, forget plot] coordinates {
(1,0.979548)
(2,0.976)
(3,0.972153)
(4,0.967561)
(5,0.962527)
(6,0.956626)
(7,0.950334)
(8,0.942727)
(9,0.933904)
(10,0.924621)
(11,0.913899)
(12,0.902986)
};

\addplot[ForestGreen,thick,only marks,mark=otimes*, mark size=2pt, forget plot] coordinates {
(1,0.999147)
(2,0.998845)
(3,0.998356)
(4,0.997707)
(5,0.996792)
(6,0.995604)
(7,0.99408)
(8,0.992058)
(9,0.988708)
(10,0.985212)
(11,0.979952)
(12,0.974286)
};

\addlegendimage{black,very thick,only marks,mark=x, mark size=3pt}
\addlegendentry{League B$\qquad$}
\addplot[red,very thick,only marks,mark=x, mark size=3pt, forget plot] coordinates {
(13,0.459412)
(14,0.45961)
(15,0.459492)
(16,0.459949)
(17,0.459171)
(18,0.459436)
(19,0.459899)
(20,0.459169)
(21,0.459544)
(22,0.46012)
(23,0.459666)
(24,0.458501)
};

\addplot[blue,very thick,only marks,mark=x, mark size=3pt, forget plot] coordinates {
(13,0.852405)
(14,0.835807)
(15,0.81763)
(16,0.797444)
(17,0.773296)
(18,0.751046)
(19,0.727298)
(20,0.702501)
(21,0.670523)
(22,0.6451)
(23,0.615997)
(24,0.588765)
};

\addplot[ForestGreen,very thick,only marks,mark=x, mark size=3pt, forget plot] coordinates {
(13,0.950322)
(14,0.937913)
(15,0.921292)
(16,0.902953)
(17,0.875784)
(18,0.849022)
(19,0.818887)
(20,0.784397)
(21,0.726905)
(22,0.684585)
(23,0.639056)
(24,0.593742)
};

\addlegendimage{black,thick,only marks,mark=diamond*, mark size=3pt}
\addlegendentry{League C$\qquad$}
\addplot[red,thick,only marks,mark=diamond*, mark size=3pt, forget plot] coordinates {
(25,0.425779)
(26,0.426269)
(27,0.426967)
(28,0.426653)
(29,0.425942)
(30,0.426393)
(31,0.426308)
(32,0.426768)
(33,0.42623)
(34,0.426667)
(35,0.426437)
(36,0.425646)
(37,0.423049)
(38,0.422939)
(39,0.42435)
};

\addplot[blue,thick,only marks,mark=diamond*, mark size=3pt, forget plot] coordinates {
(25,0.40699)
(26,0.378396)
(27,0.350842)
(28,0.322692)
(29,0.290803)
(30,0.264667)
(31,0.239731)
(32,0.217522)
(33,0.187921)
(34,0.168819)
(35,0.1501)
(36,0.133217)
(37,0.112968)
(38,0.098505)
(39,0.086034)
};

\addplot[ForestGreen,thick,only marks,mark=diamond*, mark size=3pt, forget plot] coordinates {
(25,0.402333)
(26,0.3568)
(27,0.313696)
(28,0.272029)
(29,0.220433)
(30,0.185214)
(31,0.153714)
(32,0.124772)
(33,0.085921)
(34,0.067945)
(35,0.052281)
(36,0.040339)
(37,0.02719)
(38,0.020103)
(39,0.014442)
};

\addlegendimage{black,thick,only marks,mark=star, mark size=3pt}
\addlegendentry{League D}
\addplot[red,thick,only marks,mark=star, mark size=3pt, forget plot] coordinates {
(40,0.401408)
(41,0.402016)
(42,0.401797)
(43,0.401974)
(44,0.403105)
(45,0.401715)
(46,0.402624)
(47,0.402199)
(48,0.402209)
(49,0.403116)
(50,0.401937)
(51,0.402418)
(52,0.401573)
(53,0.40253)
(54,0.402709)
(55,0.402278)
};

\addplot[blue,thick,only marks,mark=star, mark size=3pt, forget plot] coordinates {
(40,0.070693)
(41,0.060939)
(42,0.05287)
(43,0.045468)
(44,0.037269)
(45,0.031575)
(46,0.026688)
(47,0.023062)
(48,0.018127)
(49,0.01502)
(50,0.012521)
(51,0.010535)
(52,0.00847)
(53,0.006651)
(54,0.00578)
(55,0.004427)
};

\addplot[ForestGreen,thick,only marks,mark=star, mark size=3pt, forget plot] coordinates {
(40,0.021101)
(41,0.016224)
(42,0.012402)
(43,0.009623)
(44,0.005542)
(45,0.004007)
(46,0.002968)
(47,0.002035)
(48,0.001203)
(49,0.000802)
(50,0.000502)
(51,0.000359)
(52,0.000165)
(53,0.000124)
(54,0.00008)
(55,0.000046)
};
\draw (axis cs:12.5,\pgfkeysvalueof{/pgfplots/ymin})  -- (axis cs:12.5,\pgfkeysvalueof{/pgfplots/ymax});
\draw (axis cs:24.5,\pgfkeysvalueof{/pgfplots/ymin})  -- (axis cs:24.5,\pgfkeysvalueof{/pgfplots/ymax});
\draw (axis cs:39.5,\pgfkeysvalueof{/pgfplots/ymin})  -- (axis cs:39.5,\pgfkeysvalueof{/pgfplots/ymax});

\addlegendimage{line legend,very thick,red}
\addlegendentry{Difference 0$\qquad$}
\addlegendimage{line legend,very thick,blue}
\addlegendentry{Difference 10$\qquad$}
\addlegendimage{line legend,very thick,ForestGreen}
\addlegendentry{Difference 20$\qquad$}
\end{axis}
\end{tikzpicture}
\end{subfigure}

\vspace{0.5cm}
\begin{subfigure}{\textwidth}
\centering
\caption{Seeded path formation}
\label{Fair_Euro2020_Fig14b}
\begin{tikzpicture}
\begin{axis}[width = 1\textwidth, 
height = 0.6\textwidth,
xmin = 0,
xmax = 56,
ymin = -0.02,
ymax = 1.02,
ymajorgrids,
xlabel = Team rank,
xlabel style = {font =\small},
legend style = {at = {(0.5,-0.15)},anchor = north,legend columns = 4,font = \small}
]
\addlegendimage{black,thick,only marks,mark=otimes*, mark size=2pt}
\addlegendentry{League A$\qquad$}
\addplot[red,thick,only marks,mark=otimes*, mark size=2pt, forget plot] coordinates {
(1,0.473328)
(2,0.473669)
(3,0.473148)
(4,0.4735)
(5,0.472883)
(6,0.473493)
(7,0.473572)
(8,0.473371)
(9,0.4731)
(10,0.473239)
(11,0.473687)
(12,0.472798)
};

\addplot[blue,thick,only marks,mark=otimes*, mark size=2pt, forget plot] coordinates {
(1,0.983203)
(2,0.980303)
(3,0.976945)
(4,0.973271)
(5,0.968679)
(6,0.963989)
(7,0.958185)
(8,0.952076)
(9,0.944256)
(10,0.936232)
(11,0.927431)
(12,0.917962)
};

\addplot[ForestGreen,thick,only marks,mark=otimes*, mark size=2pt, forget plot] coordinates {
(1,0.999531)
(2,0.999246)
(3,0.998999)
(4,0.998571)
(5,0.99788)
(6,0.997168)
(7,0.996103)
(8,0.994766)
(9,0.992583)
(10,0.989807)
(11,0.986252)
(12,0.981972)
};

\addlegendimage{black,very thick,only marks,mark=x, mark size=3pt}
\addlegendentry{League B$\qquad$}
\addplot[red,very thick,only marks,mark=x, mark size=3pt, forget plot] coordinates {
(13,0.473454)
(14,0.473117)
(15,0.473708)
(16,0.473998)
(17,0.47336)
(18,0.473343)
(19,0.473383)
(20,0.473035)
(21,0.473179)
(22,0.473594)
(23,0.47295)
(24,0.47205)
};

\addplot[blue,very thick,only marks,mark=x, mark size=3pt, forget plot] coordinates {
(13,0.867915)
(14,0.851807)
(15,0.8335)
(16,0.815964)
(17,0.791481)
(18,0.770394)
(19,0.747783)
(20,0.72323)
(21,0.690168)
(22,0.663927)
(23,0.636157)
(24,0.609039)
};

\addplot[ForestGreen,very thick,only marks,mark=x, mark size=3pt, forget plot] coordinates {
(13,0.963616)
(14,0.953788)
(15,0.940772)
(16,0.926536)
(17,0.903125)
(18,0.88062)
(19,0.854596)
(20,0.825167)
(21,0.772489)
(22,0.733535)
(23,0.690627)
(24,0.646337)
};

\addlegendimage{black,thick,only marks,mark=diamond*, mark size=3pt}
\addlegendentry{League C$\qquad$}
\addplot[red,thick,only marks,mark=diamond*, mark size=3pt, forget plot] coordinates {
(25,0.414627)
(26,0.414725)
(27,0.415233)
(28,0.415191)
(29,0.414692)
(30,0.415283)
(31,0.415292)
(32,0.415762)
(33,0.415221)
(34,0.415499)
(35,0.415165)
(36,0.414365)
(37,0.412608)
(38,0.412399)
(39,0.413326)
};

\addplot[blue,thick,only marks,mark=diamond*, mark size=3pt, forget plot] coordinates {
(25,0.387754)
(26,0.358227)
(27,0.329609)
(28,0.301289)
(29,0.268596)
(30,0.24376)
(31,0.218852)
(32,0.198073)
(33,0.170243)
(34,0.152453)
(35,0.135532)
(36,0.120022)
(37,0.102025)
(38,0.089012)
(39,0.077402)
};

\addplot[ForestGreen,thick,only marks,mark=diamond*, mark size=3pt, forget plot] coordinates {
(25,0.364336)
(26,0.318529)
(27,0.274031)
(28,0.232381)
(29,0.177031)
(30,0.144001)
(31,0.116054)
(32,0.091673)
(33,0.063681)
(34,0.049417)
(35,0.037259)
(36,0.028074)
(37,0.019487)
(38,0.014369)
(39,0.010301)
};

\addlegendimage{black,thick,only marks,mark=star, mark size=3pt}
\addlegendentry{League D}
\addplot[red,thick,only marks,mark=star, mark size=3pt, forget plot] coordinates {
(40,0.400802)
(41,0.401296)
(42,0.401033)
(43,0.401071)
(44,0.401819)
(45,0.400883)
(46,0.402237)
(47,0.4015)
(48,0.401214)
(49,0.402162)
(50,0.400652)
(51,0.40148)
(52,0.400729)
(53,0.40149)
(54,0.401449)
(55,0.401836)
};

\addplot[blue,thick,only marks,mark=star, mark size=3pt, forget plot] coordinates {
(40,0.060158)
(41,0.051846)
(42,0.044527)
(43,0.038334)
(44,0.031383)
(45,0.02666)
(46,0.02257)
(47,0.01912)
(48,0.015283)
(49,0.012593)
(50,0.010621)
(51,0.008851)
(52,0.007133)
(53,0.005585)
(54,0.004873)
(55,0.003717)
};

\addplot[ForestGreen,thick,only marks,mark=star, mark size=3pt, forget plot] coordinates {
(40,0.010317)
(41,0.007721)
(42,0.005635)
(43,0.004124)
(44,0.002391)
(45,0.001736)
(46,0.001207)
(47,0.000834)
(48,0.000526)
(49,0.000299)
(50,0.000194)
(51,0.000156)
(52,0.000058)
(53,0.000045)
(54,0.000028)
(55,0.000019)
};
\draw (axis cs:12.5,\pgfkeysvalueof{/pgfplots/ymin})  -- (axis cs:12.5,\pgfkeysvalueof{/pgfplots/ymax});
\draw (axis cs:24.5,\pgfkeysvalueof{/pgfplots/ymin})  -- (axis cs:24.5,\pgfkeysvalueof{/pgfplots/ymax});
\draw (axis cs:39.5,\pgfkeysvalueof{/pgfplots/ymin})  -- (axis cs:39.5,\pgfkeysvalueof{/pgfplots/ymax});

\addlegendimage{line legend,very thick,red}
\addlegendentry{Difference 0$\qquad$}
\addlegendimage{line legend,very thick,blue}
\addlegendentry{Difference 10$\qquad$}
\addlegendimage{line legend,very thick,ForestGreen}
\addlegendentry{Difference 20$\qquad$}
\end{axis}
\end{tikzpicture}
\end{subfigure}

\caption{Theoretical model: qualifying probability with alternative path formations}
\label{Fig8}

\end{figure}

Figure~\ref{Fig8} reveals that changing the path formation policy can again improve fairness. Both random and seeded path formations result in a remarkable difference between Leagues B and C. As expected, seeded path formation favors less the top teams of League D than random path formation, therefore it is closer to fairness.
If all teams are equally strong ($\Delta = 0$), the qualifying probabilities of teams from Leagues A and B essentially differ only under random path formation. That is explained by the rule of hosting the semifinals in the qualifying play-offs: a play-off path usually consists of teams from the same league under regular path formation, and it usually contains one team from each league under seeded path formation, thus a team from League A has the same chance to play at home in the semifinals than a team from League B. On the other hand, random path formation does not guarantee this equality between the teams of Leagues A and B.

To conclude, our theoretical investigation reveals that unfairness is an \emph{inherent} feature of the UEFA Euro 2020 qualification and its deficiency is not only caused by the unfavorable distribution of the team's strength.

\section{Discussion} \label{Sec7}

According to Section~\ref{Sec4}, obtaining a worse position in the ranking of the national teams used for the draw of the 2018/19 UEFA Nations League can considerably increase the probability of reaching the final tournament. This feature endangers the sport's credibility and integrity as certain teams may aim to manipulate the ranking in the future.

The 55 UEFA members have been divided into the four leagues according to their UEFA national team coefficients after the conclusion of the 2018 FIFA World Cup qualifiers without the play-offs.
The coefficients are calculated as a weighted average:
\begin{itemize}
\item
20\% of the average ranking points collected in the 2014 FIFA World Cup qualifi\-cation and final;
\item
40\% of the average ranking points collected in the UEFA Euro 2016 qualification and final;
\item
40\% of the average ranking points collected in the 2018 FIFA World Cup qualifi\-cation.
\end{itemize}
The allocation of match points is explained in \citet[Annex~D]{UEFA2018e}, while \citet{Footballseeding2017} presents the details of the calculation.

The last nine matches (three matches each in the UEFA groups A, B, and H) that influence this ranking were played at the end of the 2018 FIFA World Cup qualifiers on 10 October 2017. In particular, Belgium vs.\ Cyprus was 4-0, resulting in 19,491.08 points for Cyprus, which corresponds to the 37th position. The 40th was Azerbaijan with 17,760.82 points. 
However, Cyprus would have been better off on the 40th place concerning the probability of qualification for the UEFA Euro 2020.
A conceded goal in any match means minus $500$ points, which should be divided by 10 (the number of matches in the 2018 FIFA World Cup qualification), and has a weight of 0.4 in the coefficient used for the draw of the 2018/19 UEFA Nations League. Hence, Cyprus would have been only the 40th after kicking $1740/20 = 87$ own goals against Belgium because the point difference between Azerbaijan and Cyprus was 1730.82.

While a similar tanking can be easily detected and probably sanctioned by the UEFA, a team can achieve the 40--43 positions easily if it is willing to sacrifice a whole FIFA World Cup qualification, where the probability of its success is marginal anyway.
For instance, Lithuania scored only six points in UEFA Group F during the 2018 FIFA World Cup qualification, while 18 was still not enough for the qualification. If Lithuania would have played a draw of 2-2 against Malta on 11 October 2016 instead of winning by 2-0, then it would have had $18100.74 - 1640.08 + 800.08 = 17260.74$ points, guaranteeing the 40th position in the UEFA national team coefficient ranking used for the draw of the 2018/19 UEFA Nations League \emph{ceteris paribus}. Consequently, Lithuania was severely punished for its win against Malta by the controversial format of the qualification for the UEFA Euro 2020.

\section{Conclusions} \label{Sec8}

The current paper has documented the unfairness of the qualification for the 2020 UEFA European Championship. In particular, being a top team in the lowest-ranked League D of the 2018/19 UEFA Nations League can substantially increase the probability of qualifying compared to being a bottom team in League C of the 2018/19 UEFA Nations League.
Using a more accurate prediction model to forecast match results cannot change the qualitative findings but may hide them behind the details of the statistical methodology.

Two slight modifications have been proposed to mitigate or eliminate the problem of perverse incentives.
Both require only a minor change in the path formation policy of the UEFA Euro 2020 qualifying play-offs. This rule is hidden at the deep of the regulation, is probably understood neither by the public nor by the decision makers, and contains theoretical shortcomings anyway as has already been revealed \citep{Csato2020g}.

We are afraid that the novel policy of the Union of European Football Associations (UEFA), which aims to increase the diversity of the teams playing in the UEFA Euro 2020, has been implemented without a careful analysis in advance. Our work can inspire further applications of statistical techniques that identify similar issues of unfairness in sports and propose reasonable amendments.
 
\section*{Acknowledgments}
\addcontentsline{toc}{section}{Acknowledgments}
\noindent
This paper could not have been written without \emph{my father} (also called \emph{L\'aszl\'o Csat\'o}), who has coded the simulations in Python. \\
We are grateful to \emph{Eduard Ranghiuc} for inspiration. \\
\emph{Alex Krumer}, \emph{M\'anuel M\'ag\'o} and \emph{Tam\'as Halm} gave useful comments. \\
Two anonymous reviewers provided valuable remarks and suggestions on an earlier draft. \\
We are indebted to the \href{https://en.wikipedia.org/wiki/Wikipedia_community}{Wikipedia community} for contributing to our research by summarising the tournaments discussed in the paper. \\
The research was supported by the MTA Premium Postdoctoral Research Program grant PPD2019-9/2019.

\bibliographystyle{apalike}
\bibliography{All_references}

\end{document}